\documentclass[mathleft,fleqn,%
]{an}
\usepackage{url}
\usepackage{graphicx}
\usepackage[varg]{txfonts}
\usepackage{natbib}
\bibpunct{(}{)}{;}{a}{}{,}
\begin{document}

\Pagespan{1}{}
\Yearpublication{2014}%
\Yearsubmission{2014}%
\Month{0}%
\Volume{999}%
\Issue{0}%
\DOI{asna.201400000}%

\title{Galactic Archaeology with CoRoT and APOGEE:\\
Creating mock observations from a chemodynamical model}

\author{F. Anders\inst{1,2}\fnmsep\thanks{{mailto: fanders@aip.de}}
\and C. Chiappini\inst{1,2} 
\and T. S. Rodrigues\inst{2,3,4} 
\and T. Piffl\inst{1} 
\and B. Mosser\inst{5} 
\and A. Miglio\inst{6}
\and J. Montalb\'{a}n\inst{4}
\and L. Girardi\inst{3}
\and I. Minchev\inst{1}
\and M. Valentini\inst{1}
\and M. Steinmetz\inst{1}
}
\titlerunning{Galactic Archaeology with CoRoT and APOGEE}
\authorrunning{Anders, Chiappini et al.}
\institute{
Leibniz-Institut f\"ur Astrophysik Potsdam (AIP), An der 
Sternwarte 16, 14482 Potsdam, Germany
\and 
Laborat\'orio Interinstitucional de e-Astronomia, - LIneA, 
Rua Gal. Jos\'e Cristino 77, Rio de Janeiro, RJ - 20921-400, Brazil
\and 
Osservatorio Astronomico di Padova -- INAF, Vicolo 
dell'Osservatorio 5, I-35122 Padova, Italy
\and
Dipartimento di Fisica e Astronomia, Universit\`a di Padova, Vicolo dell'Osservatorio 2, I-35122 Padova, Italy
\and 
LESIA, Observatoire de Paris, PSL Research University, CNRS, Universit\'{e} Pierre et Marie Curie, Universit\'{e} Denis Diderot, 92195 Meudon, France
\and
School of Physics and Astronomy, University of Birmingham, 
Edgbaston, Birmingham, B15 2TT, United Kingdom
}

\received{XXXX}
\accepted{XXXX}
\publonline{XXXX}


\abstract{
In a companion paper, we have presented the combined asteroseismic-spectroscopic dataset obtained from CoRoT lightcurves and APOGEE infra-red spectra for 678 solar-like oscillating red giants in two fields of the Galactic disc (CoRoGEE). We have measured chemical abundance patterns, distances, and ages of these field stars which are spread over a large radial range of the Milky Way's disc. 
Here we show how to simulate this dataset using a chemodynamical Galaxy model. We also demonstrate how the observation procedure influences the accuracy of our estimated ages.
}

\maketitle

\section{Introduction}

Galactic models make predictions for the distribution of stars and gas in the multi-dimensional space consisting of time, kinematics and chemical composition.
Therefore, one of the basic problems of Galactic Archaeology -- the science of inferring the current state and the history of the Milky Way from present-day observations (e.g., \citealt{Pagel1997, Freeman2002}) -- is dimensionality reduction. For a given dataset, we are looking for the most robust and telling statistical relations to constrain these models.

Asteroseismology of red giants delivers new promising constraints to Milky Way models since it provides masses and ages of distant field stars with unprecedented precision (e.g., \citealt{Miglio2013a}).
The present work and an accompanying series of papers (\citealt{Chiappini2015}; Anders et al., subm. to A\&A) explore the power of asteroseismic constraints in Galactic Archaeology: we present one of the first attempts to combine stellar physics, asteroseismology, statistics, and spectroscopy -- to learn about the chemo-dynamical history of our Galaxy. Specifically, we combine data from the infrared spectroscopic stellar survey APOGEE \citep{Majewski2015} with asteroseismic data from the CoRoT mission \citep{Baglin2006}. 
In this paper, we describe how we simulated mock CoRoT-APOGEE (CoRoGEE) observations of the chemodynamical N-body Galaxy model of \citet*[][MCM]{Minchev2013, Minchev2014b}. 

\begin{figure*}
\centering
 \includegraphics[trim=0cm 0cm 0cm 0cm, clip=true, width=.8\textwidth]{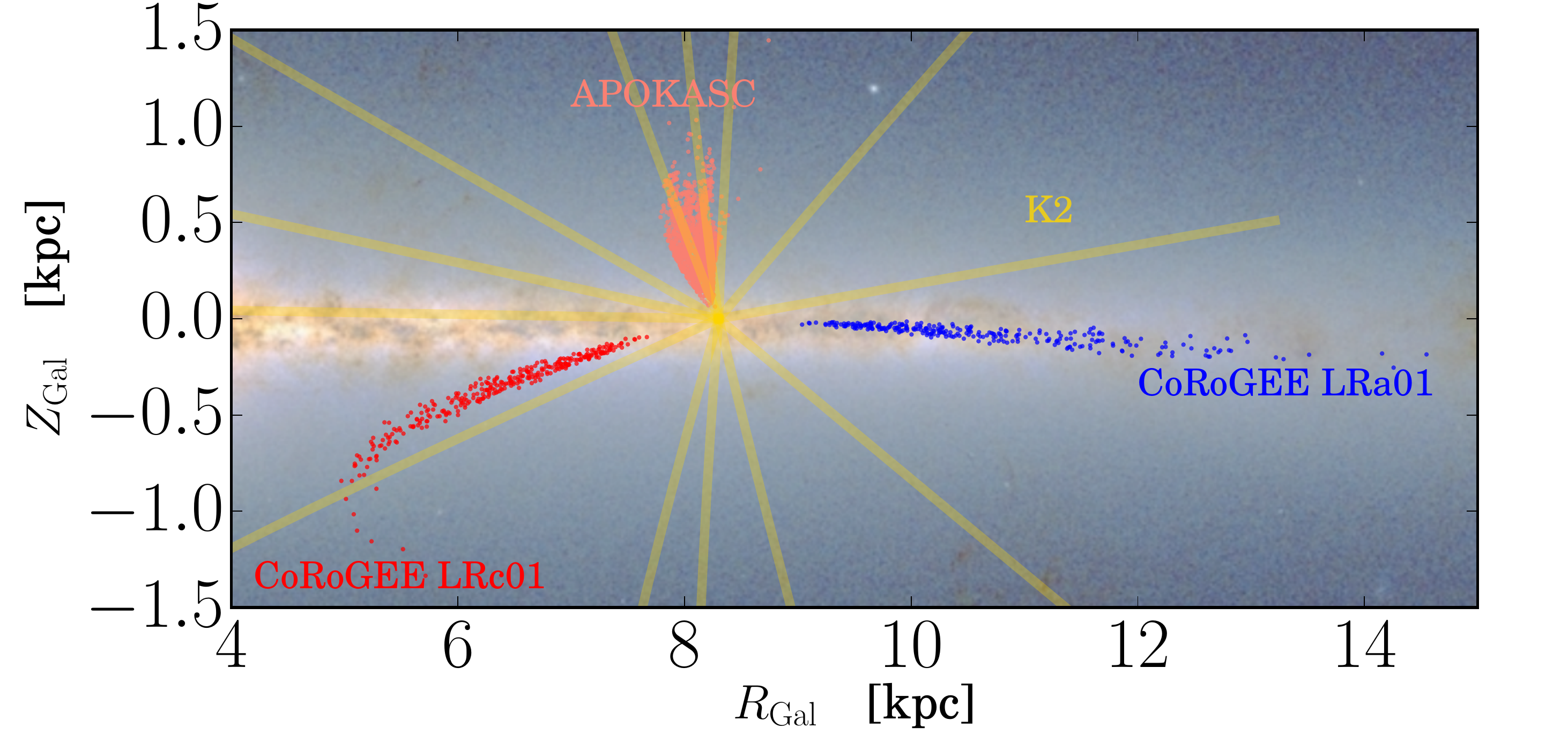}\\
\caption{\small Location of the APOGEE samples with seismic and spectroscopic observations in Galactocentric cylindrical coordinates. The K2 mission and its spectroscopic follow-up campaigns are presently adding several new sightlines to this picture ({\it yellow rays}).}
\label{rzplane}
\end{figure*}

\section{The dataset}

We have assembled a comprehensive dataset (stellar parameters, elemental abundances, kinematics) of more than 600 solar-like oscillating red giant stars which have been observed by both CoRoT and APOGEE (CoRoGEE). Table 1 gives an overview of the dataset; Fig. \ref{rzplane} shows the distribution of our stars in Galactocentric cylindrical coordinates. The details of our analysis are provided in Anders et al. (A\&A, subm.).

\begin{table}
\caption{Summary of the available CoRoGEE data.}
\begin{tabular}{ l l }
\hline
\hline
CoRoT-APOGEE stars & 690\\
\hline
\quad  with ``good'' spectroscopic parameters & 678 \\
\quad  \& ``good'' asteroseismic parameters & 664 \\
\quad  \& $|\log g_{\mathrm{ASPCAP}} - \log g_{\mathrm{seismo}}|<0.5$ & 617 \\
\hline
Converged stellar parameters \& distances & 606 \\
\quad Field LRa01 (outer disc) & 282 \\
\quad Field LRc01 (inner disc) & 326 \\
\hline
\end{tabular} 
\end{table}

Using an updated version of the Bayesian stellar parameter estimation code PARAM \citep{daSilva2006}, we have determined the radii, masses, ages, and distances of the CoRoGEE stars by comparing the measured spectroscopic effective temperatures, metallicities, and asteroseismically determined $\Delta\nu$ and $\nu_{\mathrm{max}}$ with stellar evolutionary models. We achieve typical precicions of $\sim3\%$ in radius, $\sim9\%$ in mass, and $\sim25\%$ in age. By combining our stellar radii measurements with multi-wavelength photometry, we also derive very precise distances (precise to $\sim2\%$) and extinctions. The details are described in \citet{Rodrigues2014}.



The first result obtained with the CoRoT-APOGEE dataset was the discovery of a population of disc stars which do not follow the relation between the [$\alpha$/Fe] abundance ratio and age predicted by canonical chemical evolution models of the Galactic disc. In \citet{Chiappini2015}, we discuss several scenarios that can be invoked to explain the existence of these objects, and the fact that these stars are much more prevalent in the inner CoRoT field. No conclusive explanation has been presented so far, but possible solutions involve stellar mimicry (old stars disguised as younger ones because of close-binary evolution \citep{Yong2016} or stellar mergers), abundance anomalies in star-forming bubbles, and a peculiar chemical evolution near the corotation radius of the Galactic bar.

\section{CoRoGEE mock samples from a chemo-dynamical model}\label{mockapp}

\begin{figure}\centering
\includegraphics[trim=0cm 6cm 0cm 0cm, clip=true, width=.46\textwidth]{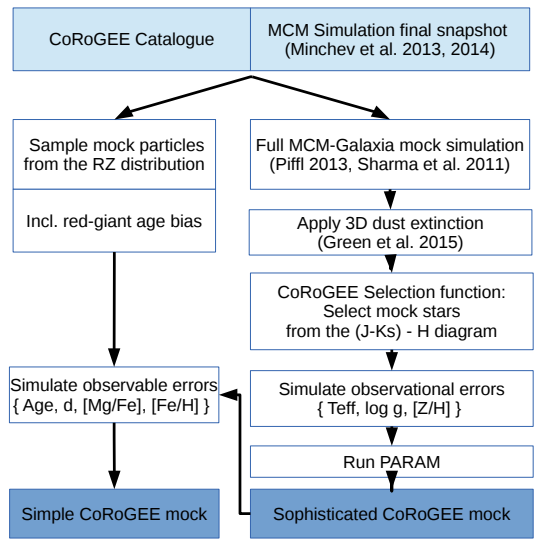}
 \caption{Scheme illustrating how our two versions of the CoRoGEE mock observations were obtained from the MCM model.}
 \label{mockchart}
\end{figure}

The direct interpretation of astronomical survey data is often hampered by non-trivial selection effects. As pointed out in, e.g., \citet{Binney2015}, the comparison of survey catalogues with a Galactic model is much easier when a mock observation of the model is created. 

In this Section we describe how to select a CoRoGEE-like sample from an N-body simulation, using the example of the 
MCM model \citep{Minchev2013, Minchev2014b}. We have chosen two different paths to simulating the observations: a straightforward ``simple'' mock, and a more sophisticated one which uses a mock observation tool \citep{Piffl2013} based on the Galaxia stellar population synthesis code \citep{Sharma2011}. The procedures leading to the two versions of mock observations are sketched in Fig. \ref{mockchart}.

\begin{figure*}\centering
\includegraphics[width=.395\textwidth]{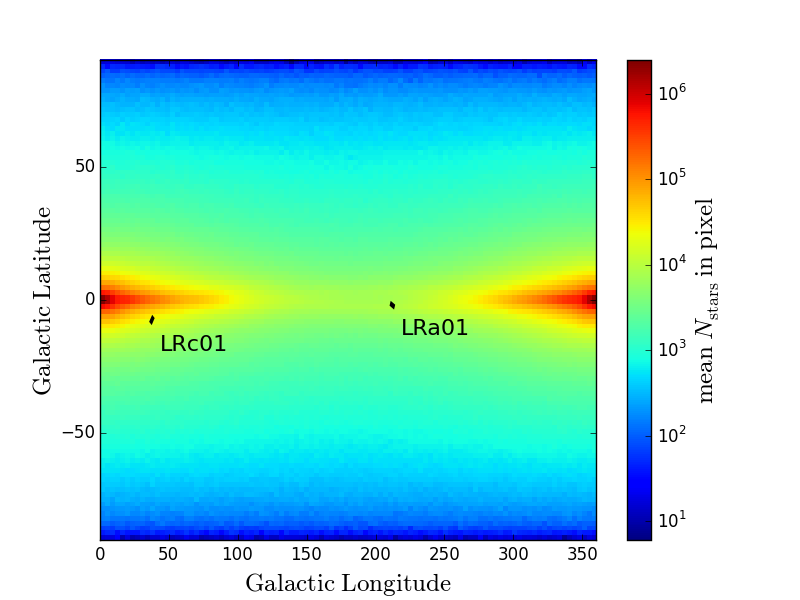}
\includegraphics[width=.59\textwidth]{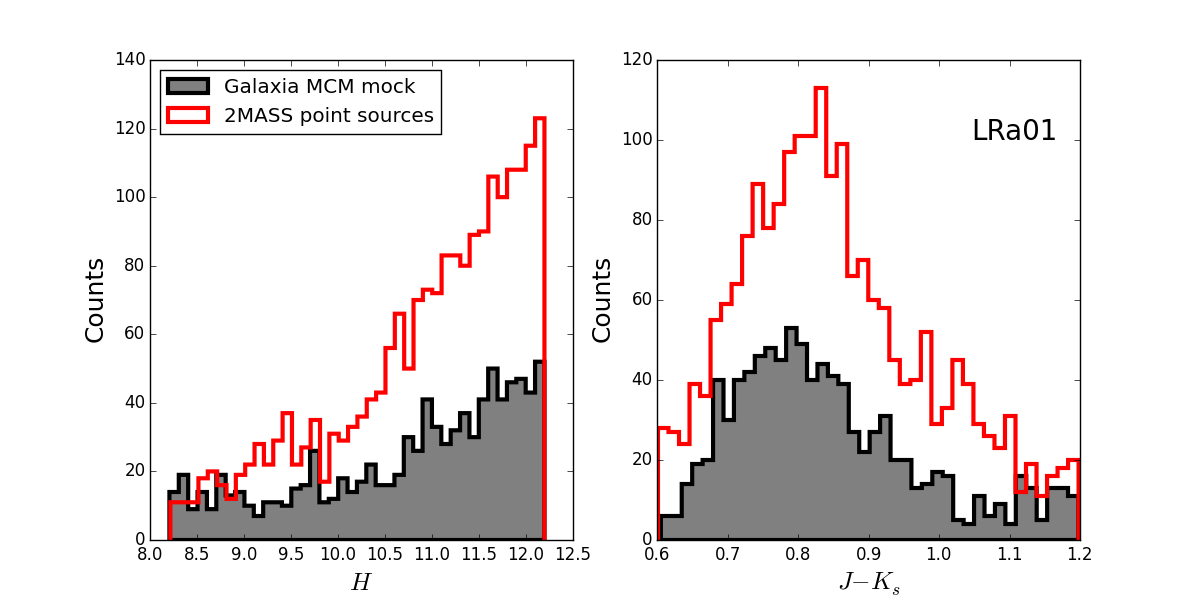}\\
\includegraphics[width=.395\textwidth]{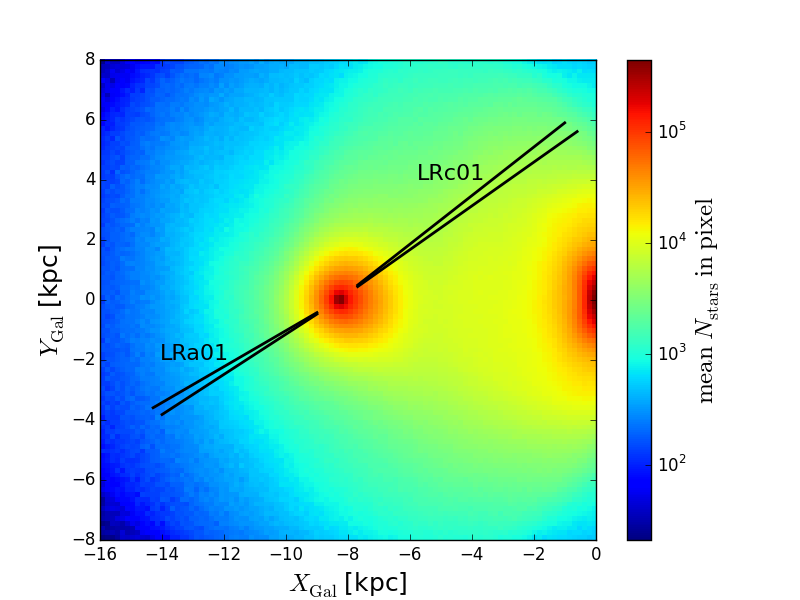}
\includegraphics[width=.59\textwidth]{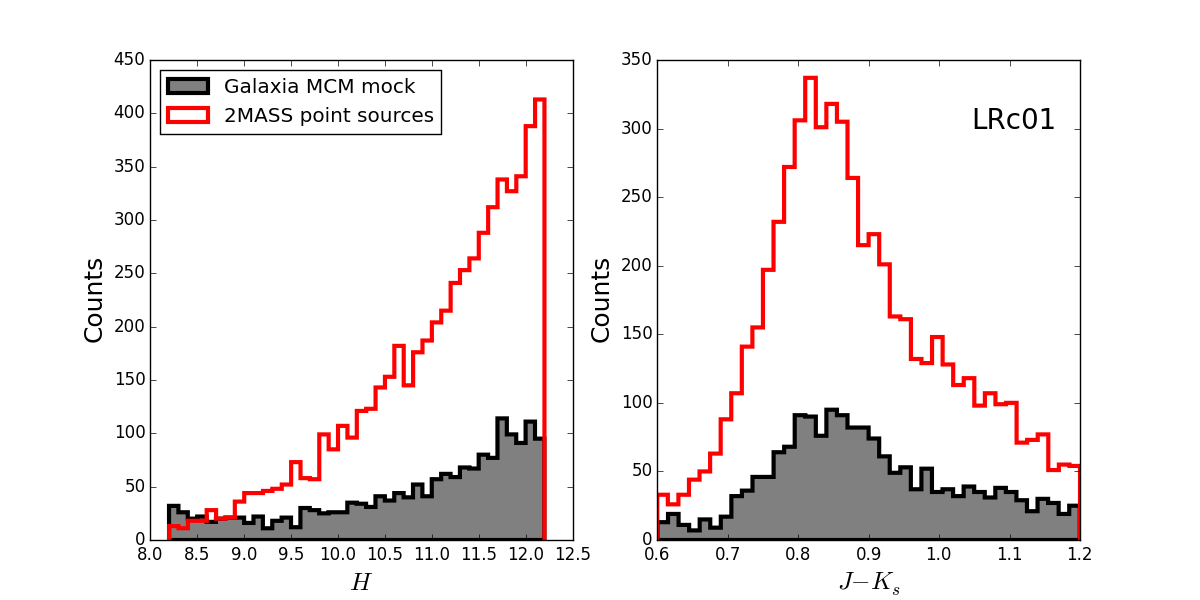}
 \caption{Star counts in the MCM Galaxy mock. {\it Left:} Density distribution of all simulated MCM stars (magnitude limit $H_0=13.0$) in $(l, b$; top left) and $(X_{\rm Gal}, Y_{\rm Gal}$; bottom left). {\it Middle and right:} $H$ magnitude and $(J-K_s)$ star counts in the two CoRoT fields, comparing 2MASS (red histograms; \citealt{Cutri2003}) and the MCM mock Galaxy (grey histograms). }
 \label{cmd_distr}
\end{figure*}

 \begin{figure*}\centering
\includegraphics[trim=0cm 0cm 4.8cm 0cm, clip=true, width=.43\textwidth]
{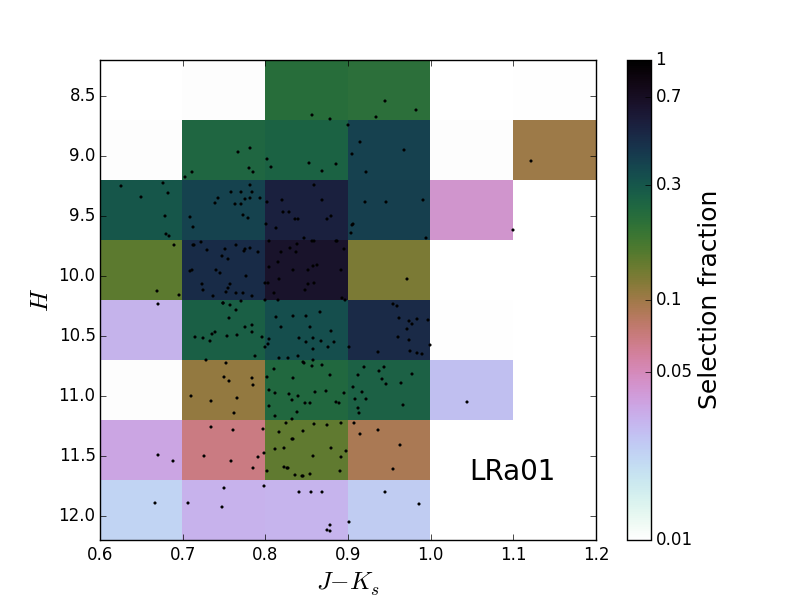}
\includegraphics[trim=1.3cm 0cm 0cm 0cm, clip=true, width=.525\textwidth]
{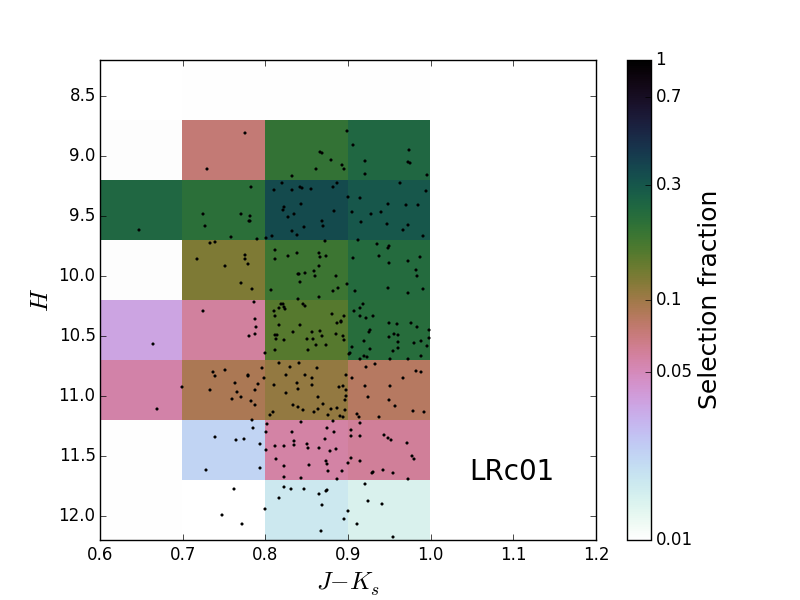}
 \caption{$H$ vs. $J-K_s$ colour-magnitude diagram (CMD) for the two CoRoGEE fields. The colour in each CMD box shows the selection fraction ($N_{\rm CoRoGEE}/N_{\rm 2MASS}$) in this box. We used the same boxes to simulate the CoRoGEE selection function for the ``sophisticated MCM mock''.}
 \label{cmds}
\end{figure*}

\subsection{Sophisticated mock}\label{soph}

The original Galaxia population synthesis code \citep{Sharma2011} uses the analytic Besan\c{c}on Milky Way model \citep{Robin2003} and creates synthetic Galactic stellar populations in a given part of the sky. Additionally, it allows the user to include a stellar halo from an N-body simulation, i.e. a model in which the kinematic distribution functions are not analytic any more, but are taken from the mass particle distributions of the input simulation. \citet{Piffl2013}\footnote{\url{https://publishup.uni-potsdam.de/files/6790/piffl_diss.pdf}} generalised this idea and first used the MCM model as an input for the Galaxia code in the context of a simulated RAVE survey. By spawning mock stars from the MCM mass particles (each star inherits its age and chemical properties from the parent particle) he showed that the model could recover realistic correlations between the kinematics and the chemical abundances of the stars, while it was not possible to obtain an absolute match with star counts and global kinematic parameters of the Milky Way. Here, we use the same code to simulate a CoRoGEE-like sample from the MCM galaxy.

We first simulated a complete synthetic photometric all-sky survey from the solar position\footnote{We assume $R_{\rm Gal, \odot}=8.3$ kpc, in line with recent estimates (see e.g., \citealt{Bland-Hawthorn2016})} up to a limiting magnitude of $H_0=13$ from the MCM galaxy using the modified Galaxia code \citep{Piffl2013}. This translates the $9.5\cdot10^5$ input N-body particles into $7.8\cdot10^7$ mock stars (see density maps in Fig. \ref{cmd_distr}). We then calculated observed magnitudes for the mock stars in the CoRoT fields using the new PanSTARRS-1 3D extinction map of \citet{Green2015}\footnote{\url{http://argonaut.skymaps.info/}}. The resulting colour and magnitude distributions up to the magnitude limit of CoRoGEE ($H=12.2$) are also shown in Fig. \ref{cmd_distr}. As expected, the absolute star counts are not well matched by the MCM-Galaxia model, but the relative distributions in the colour-magnitude diagram (CMD) are reproduced (see \citealt{Piffl2013} for a discussion). 
In the next step, we applied the effective CoRoGEE selection function (assuming that it only depends on $H$ and $J-K_s$) by randomly selecting the observed number of stars from small boxes in the CMD (see Fig. \ref{cmds}).\footnote{The justification for this approximation of the CoRoGEE selection is given in Anders et al. (2016, subm. to A\&A).} We further simulated Gaussian observational errors in the spectroscopic stellar parameters {$T_{\rm eff}, \log g,$ [Z/H]} and magnitudes, and then ran the Bayesian parameter estimation code PARAM \citep{Rodrigues2014} to recover measured masses, radii, and ages. 

\subsection{Simple mock}\label{simple}

A simpler way to simulate a CoRoGEE sample from the MCM simulation is to randomly select the most representative MCM particles from their distribution in configuration space. However, when we put the Sun at the correct distance to the Galactic center, the number of available particles is too small to yield enough mock stars in the two CoRoT fields.
Therefore, we smoothed over the azimuthal angle in the Galactocentric cylindrical frame, and drew the mock stars directly from the observed distribution in the $R_{\mathrm{Gal}}-Z_{\mathrm{Gal}}$ plane. Because red giant stars do not sample all ages evenly, we simulated this red-giant age bias by assuming that a red giant of age $\tau$ is picked with a probability $\propto (\tau+1\,\mathrm{Gyr})^{-0.7}$.\footnote{From population synthesis modelling with TRILEGAL, we find that this bias depends very weakly on the position in the Galaxy. It is also consistent with the age bias that \citet{Casagrande2016} determined for the {\it Kepler} field with different methods (their Fig. 12d).} 

Finally, we added typical observational errors in age, distance, and metallicity. While the distance and metallicity uncertainties could be assumed to be small and Gaussian ($\sim 2\%$ and 0.04 dex, respectively), the statistical age errors are not straightforward to simulate. We therefore opted for the following data-driven approach: we use the sophisticated mock to estimate the age errors. For each star, a random age error was added according to the relative error distribution shown in Fig. \ref{mockages}. 

\subsection{Simulated age distributions}

\begin{figure}\centering
\includegraphics[width=.46\textwidth]{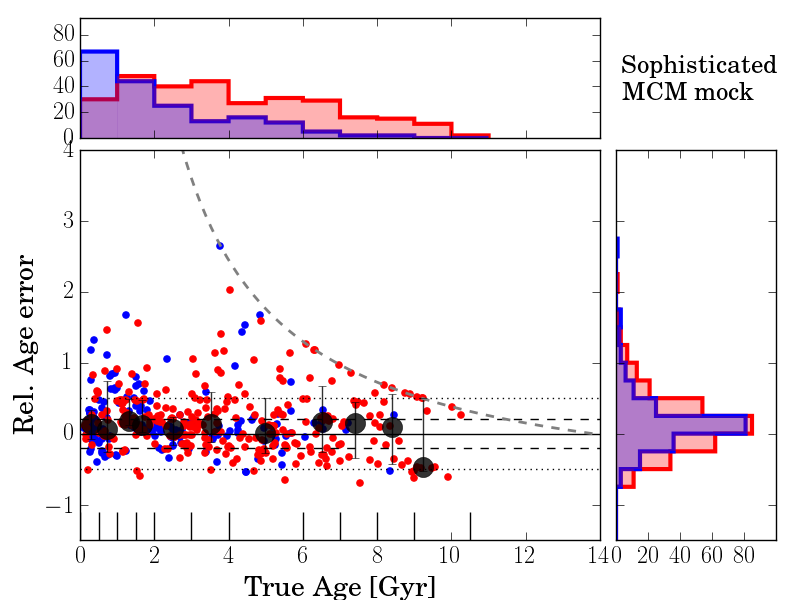}
 \caption{Relative PARAM age errors $\frac{\tau_{\rm PARAM} - \tau_{\rm true}}{\tau_{\rm true}}$ for the sophisticated version of the MCM-CoRoGEE mock as a function of the true age of the parent N-body particle.
 The simulated stars in LRa01 and LRc01 are shown in blue and red, respectively. The black symbols correspond to the median age error in each age bin indicated on the x-axis.
 The various lines correspond to a one-to-one relation, 20\% and 50\% deviation, and the age boundary at 13.7 Gyr. 
 }
 \label{mockages}
\end{figure}

Fig. \ref{mockages} shows how well our method is able to recover stellar ages, using the sophisticated MCM mock described above. It is evident that our individual age estimates should be used with caution, in particular for measured ages $>4$ Gyr. However, we confirm that a small measured age does correspond to a true small age in almost all cases, thus strengthening the conclusions of \citet{Chiappini2015}.
More details about statistical and systematic uncertainties involved in our age determinations are presented in Anders et al. (2016). 

\begin{figure}\centering
\includegraphics[width=.46\textwidth]{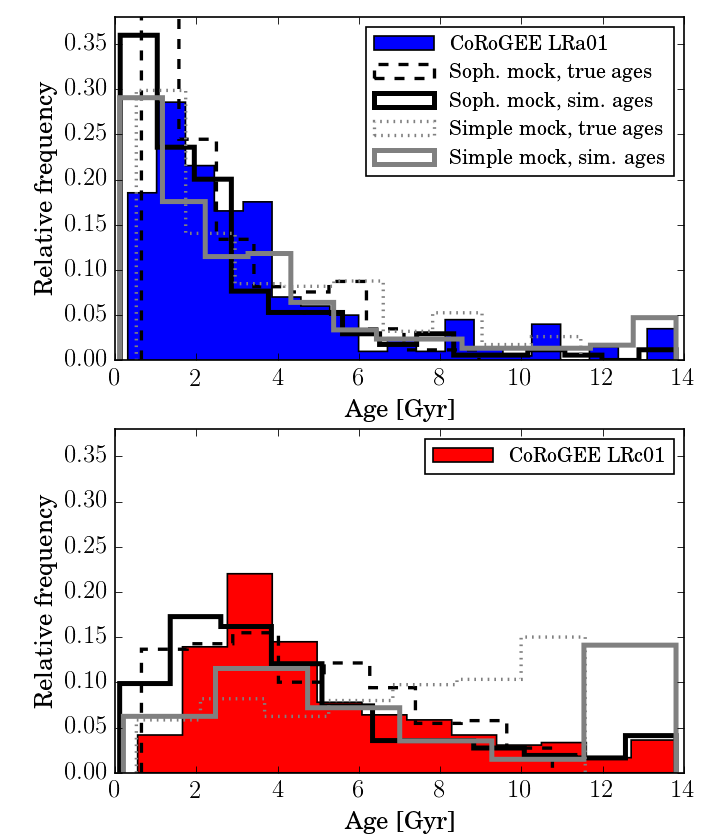}
 \caption{Simulated and recovered age distributions for the two CoRoGEE fields LRa01 (top) and LRc01 (bottom).}
 \label{mockages2}
\end{figure}

In Fig. \ref{mockages2}, we take a first look at the simulated ``true'' age distributions in the two CoRoT fields (grey histograms), the effect of adding age errors on this distribution (black histograms), and compare these with the measured age distributions of the real data (filled histograms).

While the simulated age distributions of the sophisticated mock match the data surprisingly well in LRc01, we see striking differences in the relative number of old stars in LRa01. 
Conversely, the simple mock performs better for LRa01, while it overpredicts the number of old stars in LRc01. We suggest that this may be related to a) a more complex selection function, or b) a stronger age bias towards the inner Milky Way.

\section{Summary}

In our companion paper (Anders et al. 2016), we demonstrate, in line with previous works, that combining seismology and spectroscopy brings us one step further in obtaining meaningful ages of field stars. We also show that our sample can be used to formulate new chemodynamical constraints on the evolution of the Milky Way disc over a large range in Galactocentric distance and ages.

The simulations presented in this paper have shown that some notes of caution are due: we demonstrated that the absolute age scale of our isochrone ages is prone to systematic shifts. We also remind the data user to be very careful when interpreting small subsets of the data, and to refrain from interpreting single data points. 

In follow-up works we will explore the individual-element abundance space opened by APOGEE and provide a detailed comparison with a (semi-)cosmological chemodynamical N-body simulation, using mock observation tools.
One of the key questions of Galactic Archaeology which our sample should help to answer is constraining the migration efficiency in the Galactic disc as a function of time and position.

\acknowledgements
We thank M. Schultheis for useful comments. 
TSR acknowledges support from CNPq-Brazil. BM acknowledges financial support from the ANR program IDEE Interaction Des \'Etoiles et des Exoplan\`etes.
The CoRoT space mission, launched on December 27 2006, was developed and 
operated by CNES, with the contribution of Austria, Belgium, Brazil, ESA (RSSD 
and Science Program), Germany and Spain.
This research has made use of the ExoDat Database, operated at LAM-OAMP, 
Marseille, France, on behalf of the CoRoT/Exoplanet program.\\

Funding for the SDSS-III Brazilian Participation Group was provided by the 
Ministério de Ciência e Tecnologia (MCT), Funda\c{c}\~{a}o Carlos Chagas Filho 
de Amparo à Pesquisa do Estado do Rio de Janeiro (FAPERJ), Conselho Nacional de 
Desenvolvimento Científico e Tecnológico (CNPq), and Financiadora de Estudos e 
Projetos (FINEP).\\
Funding for SDSS-III was provided by the Alfred P. Sloan Foundation, the 
Participating Institutions, the National Science Foundation, and the U.S. 
Department of Energy Office of Science. The SDSS-III web site is 
\texttt{http://www.sdss3.org/}.\\
SDSS-III was managed by the Astrophysical Research Consortium for the 
Participating Institutions of the SDSS-III Collaboration including the 
University of Arizona, the Brazilian Participation Group, Brookhaven National 
Laboratory, Carnegie Mellon University, University of Florida, the French 
Participation Group, the German Participation Group, Harvard University, the 
Instituto de Astrofisica de Canarias, the Michigan State/Notre Dame/JINA 
Participation Group, Johns Hopkins University, Lawrence Berkeley National 
Laboratory, Max Planck Institute for Astrophysics, Max Planck Institute for 
Extraterrestrial Physics, New Mexico State University, New York University, 
Ohio State University, Pennsylvania State University, University of Portsmouth, 
Princeton University, the Spanish Participation Group, University of Tokyo, 
University of Utah, Vanderbilt University, University of Virginia, University 
of Washington, and Yale University.

\bibliographystyle{an}
\bibliography{FA_library}

\end{document}